# Ranking by Momentum based on Pareto ordering of entities

Tomasz Imielinski

Articker.org

**Abstract:**

Seemingly simple questions can be surprisingly difficult. A truly simple question is 'What books did Charles Dickens write?' that can be definitively answered in milliseconds from structured data. A *seemingly* simple question that is actually *surprisingly* difficult is 'What are the fastest-rising books being sold?'. The difficulty lies in formalizing the definition of Momentum that does not unintentionally bias one set (e.g. smaller authors have highest growth rate) over another (e.g. best selling authors have highest absolute volume).

Thus, we seek to create a definition of Ranking by Momentum that addresses the challenges of providing a satisfying answer to this complex challenge. To state the problem: given a set of changing entities, which ones are the most uptrending over some time T? Which entities are standing out as the biggest "movers"?

To answer this question we define the concept of *momentum*. Two parameters – *absolute gain* and *relative gain* over time T play the key role in defining momentum. Neither alone is sufficient, since they are each biased towards a subset of entities. In the above 'fastest-selling author' example, absolute gain favors large entities (sales rose from 100,000 to 150,000 units), while relative gain favors small entities (sales rose from 10 to 200 units).

To accommodate both absolute and relative gain in an unbiased way, we define Pareto ordering between entities. For entity E to dominate another entity F in Pareto ordering, E's absolute and relative gains over time T must be higher than F's absolute and relative gains respectively. Momentum leaders are defined as maximal elements of this partial order – the Pareto frontier. We show how to compute momentum leaders and propose linear ordering among them to help rank entities with the most momentum on the top.

Additionally, we show that when vectors follow power-law distribution then we achieve a new property which we call the *Small Pareto Frontier* property. The size of the set of momentum leaders (Pareto frontier) is of the order of $O(log(n)+1)^2)$, thus it is very small.

## 1. Introduction

Ranking by momentum has wide applications in such diverse fields as commerce, research, and consumer entertainment. What are the fastest-growing products in inventory? Which group of employees has the most dramatic change in attrition? What YouTube videos have highest momentum? What stocks, artists, or researchers are showing the most interesting rise in citations? It is a form of a summary of change – who is driving it?

Our paper has been motivated by the Articker platform [1], where we are following over 100,000 artists over time measuring their media presence continuously. In this example, every month we answer the question 'Which artists carry the most momentum?' on the website www.articker.org .

To answer this question we need to define the concept of momentum formally.

There are two basic ways of measuring a change of dynamically evolving entities: *absolute change and relative change*. A large entity may display substantial absolute gain, which is still small in relative terms, due to the large base value of such an entity. A small entity may have quite a small absolute gain, despite relative gain is very large. Which entity carries more momentum, the one with large absolute gain, or one with large relative gain? It all depends on how large the entity is in the first place.

To accommodate both absolute and relative gain in an unbiased way, we define partial order between entities. For one entity, E, to dominate another entity F in this new partial order, both absolute and relative gains over T of E, have to be higher than F's absolute and relative gains. Technically it is *Pareto ordering* of two-dimensional vectors cite[1]. Momentum leaders are defined as maximal elements of this partial order, which is the Pareto frontier for the two-dimensional vectors <absolute gain, relative gain>. We show how to compute momentum leaders and propose linear ordering among them to help rank entities with the most momentum on the top.

### Example 1 (Videos)

Let V1 and V2 be two YouTube videos, one with 1 billion views and another with 1 million views, as measured on May 1, 2021.

Let's assume that V1 gains 10 million views and V2 gains 1 million views over, say, the next three months. Which of the two carries more momentum: V1 or V2?

One is tempted to say, that it is V2, after all, it *doubles* its views over the 3 months, from 1M to 2M. while V1 only increases their views by a meager 1%. On the other hand, V1 gains 10M over three months, 10 times as many as V2. It is a toss-up! V2 dominates V1 in relative gain and V1 dominates V2 in absolute gain over 3 months.

Relative gain alone cannot be a measure of momentum, otherwise, a video V3 with just 100 original views on May 1, which increased its views over the next 3 months to 1000 (10 fold!), would carry even more momentum than V2.

Generally relative gain alone favors smaller entities (videos with very few views), while absolute gain, favors larger entities – very popular videos. Thus, momentum is a two-dimensional measure.

None of the three videos V1, V2, V3 is comparable to the remaining two. This is the case because none of the three dominates the remaining ones *both* in absolute and relative gain.

Let's build on this example and consider the following 8 videos, in Table 1, sorted by a total number of views. Additionally, we show views gained last week and the relative increase of views over the last week.

| Video | Total Views Till Last Week | Views Last Week | Relative gain |
|---|---|---|---|
| V1 | 10B | 50M | 0.50% |
| V2 | 1B | 40M | 4.17% |
| V3 | 500B | 80M | 19.10% |
| V4 | 500B | 50M | 11.10% |
| V5 | 300B | 100M | 50.00% |
| V6 | 200B | 50M | 33.00% |
| V7 | 100B | 50M | 100.00% |
| V8 | 89B | 72M | 1000.00% |

**Table 1.** *Example: 8 videos sorted by a total number of views till last week with absolute weekly and relative weekly gain of views.*

Which videos lead the momentum? One approach would be to name V3, as gaining the highest number of views last week (80M), and V8 with the highest relative gain over last week, 1000%.

But how about V5?

It dominates V8 in terms of last week's views (100K to 72K) and it dominates V3 in terms of relative gain (50% to 11.1%).

No video would have **both** higher absolute last week views as well as highest relative gain over last week than V5, just as no such video exists for V3 and V8.

Thus, we postulate that V3, V5, and V8 are momentum leaders. Indeed, all other videos are dominated by one of these three – i.e. both their absolute gain and relative gain are smaller than one of the V3, V5, and V8. Indeed V1 is dominated by V3, so is V2. V4, V6, V7 are all dominated by V8. V5 stands by itself. In other words, V3, V5, and V8 form the Pareto frontier for the {V1…V8}.

Imagine that we considered two more videos here V9 with 250K total views and 80K views last week (relative gain 47%) V10 with 200M total views and 25M views added last week (relative gain of 14.3%). It is easy to see that V9 and V10 would also be momentum leaders – since no video would dominate V9 or V10 both in relative and absolute gain. Moreover, neither V9 nor V10 can be improved anymore. Thus, both V9 and V10 would join the Pareto frontier – the set of Momentum leaders.

### Example 2 (Investment Positions)

Suppose we examine all investment positions of our portfolio of investments. Each position is defined by funds invested in a stock. There are two ways to evaluate a position: the total absolute return of a position in dollars, and yield, or relative return. of this position in %.

We may have large positions with sizable total returns, which yield small relative returns. On the other hand, some small positions may be enjoying enormous yield. For example, we may have invested $100,000 in Bank of America with a total yield of 10%, and also invested $2000 in Tesla with a yield of 400%. Which position performed better? Bank of America achieved a higher return, because of the size of our position, but surely the Tesla position is the one we would brag about at a party.

Too bad we did not put more money in Tesla!

Unfortunately, as often happens, when large risk is involved, we tend to invest less and even if the risk paid off, absolute return is not that impressive.

Ranking investment positions are similar to the momentum ranking. It is a two-dimensional problem.

Neither absolute gain nor relative gain is sufficient to characterize momentum. Both of them matter. This is the approach proposed in this paper. Our approach applies to a wide range of applications such as

i. Ranking YouTube videos by momentum are dependent on the absolute and relative increase of views over some time
ii. Ranking Twitter accounts by momentum is dependent on the absolute or relative increase of the number of followers over some time
iii. Ranking of investment positions depending on absolute return and overall yield of the position.
iv. The ranking net worth of individuals by momentum is dependent on the absolute or relative increase of their net worth over some time
v. Ranking book sales by momentum is dependent on the absolute or relative increase of their sales over a period of time etc.

## 2. Basic Notions

Let E be a linearly ranked set of N entities and T be a period of time. We will refer to the entities by their ordinals ranging from 1,...N, where 1 refers to the entity ranked first and N to the one ranked last. Ranking can be based on some underlying scoring function with entities ranked in descending order of the score. But the absolute value of the score is irrelevant here – just the ranking it imposes. For example, YouTube videos can be ranked by the total number of views, Stocks by market capitalization, etc.

Let g: <1...N> -> R be a function called absolute gain (over period T). Let r be another function from <1...N> ->R called relative gain (over period T).

The triple

E*=<N, g, r>

Is called a Δ-system.

Δ-system represent a single snapshot of change of all entities from 1 to N, over the same period of time T. The period of time T can range from seconds to months. The absolute gain and relative gain functions refer to changes of entity scores over T. Absolute gain g refers to the absolute gain of score over period T. Relative gain, r refers to percentile points – comparison of absolute gain to the score of the entity before the period T). For example, the absolute gain of a YouTube video can correspond to the additional views it gained over the last 24 hours, while relative gain can be the ratio of absolute gain and the total number of views for this video.

Given two entities e, f, we define

e <<$_m$ f  iff   g(e) < g(f) and r(e) <  r(f)

Clearly <<$_m$ is Pareto ordering, the partial order, which is not total. That is, there are incomparable pairs of entities e and f, such that neither

e <<$_m$ f nor  f <<$_m$ e is true

By a *momentum leader* in the Δ-system <N,g,r>, we mean a maximal element in <<$_m$.

This is illustrated by Figure 1 below

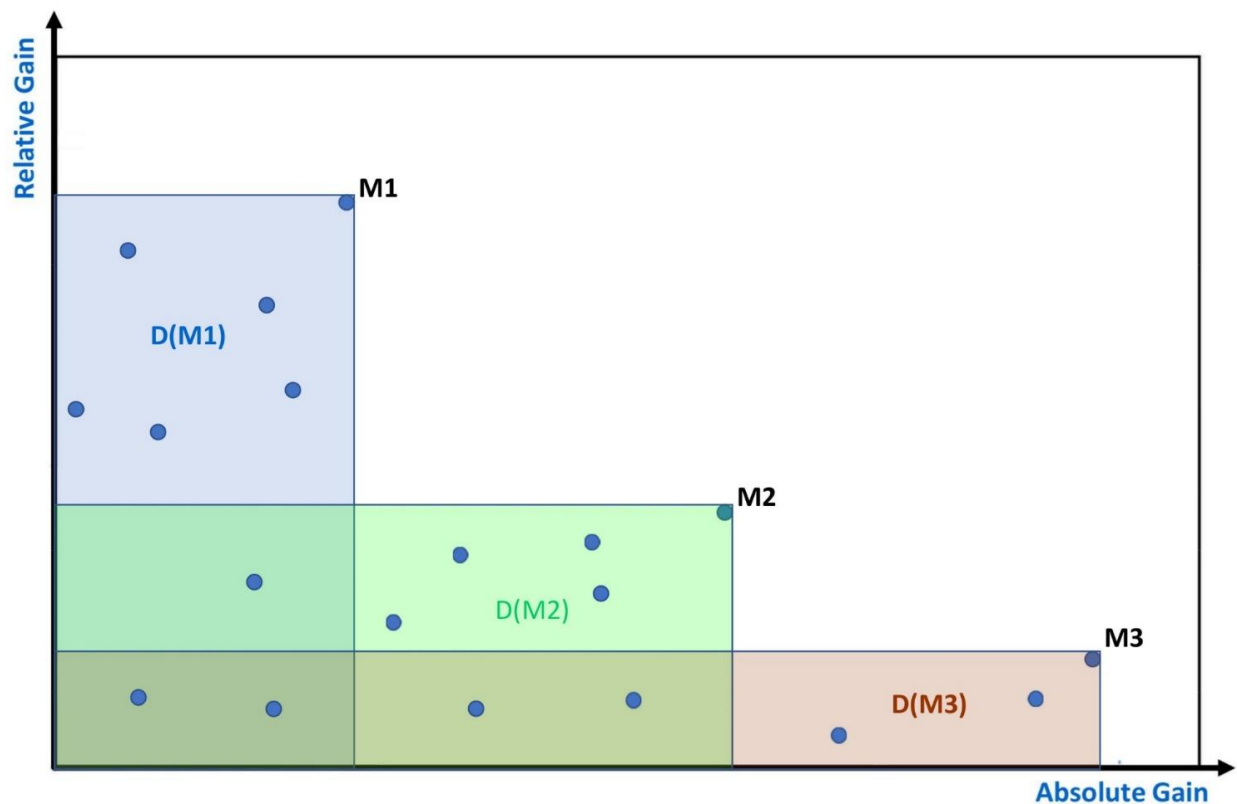

**Figure 1.** *Momentum leaders (Pareto Frontier) M1, M2, and M3 dominate areas in their respective rectangles.*

Notice that M1, M2, and M3 in Figure 1 are incomparable. However, each of M1, M2, and M3 carries more momentum than unlabeled entities in their corresponding rectangles. D(Mi) is defined as the set of all entities dominated by the momentum leader Mi.

*Momentum ordering is another application of Pareto ordering and Momentum leaders are defined as Pareto frontier for $<<_m$.*

### 3. Finding momentum leaders.

We can construct the Directed Acyclic Graph, $G(<<_m)$ in linear time by scanning all entities in the ∆-system

<N, g, r>.

We can also establish the set of momentum leaders as maximal elements in $G(<<_m)$, in one data scan.

Indeed, we can build binary table M[ , ] such that <e1, e2> is in M iff $e1 <<_m e2$ :

*{M:=NIL*

*For i=1…n*

  *Begin*

   *for $e_j <<_m e_i$*

   *M:=M union {<$e_j, e_i$>}*

  *End}*

The maximal elements of Graph $G(<<_m)$ are determined from M as entities that do not appear in the first column of M.

Momentum leaders can be thought of best summaries of change. They "lead the pack" in the collective change of a set of entities.

### Related Work

Pareto ordering, Pareto dominance and Pareto frontier have been widely used in science and engineering [8], particularly in Finance and Economics for many decades [2], [3], [4], [7]. The problem of maxima of point set in multidimensional space has been considered in [4] and [5]. Different notion of Momentum is widely used in finance – referring to the gradients of price movement of securities.

Our approach is first to our knowledge which applies Pareto ordering to discover the momentum leaders across different disciplines with changing set of entities. We also identify for the first time the small frontier property associated with the power law distribution of absolute and relative gains (section 3.2)

### Example: Top Trending music videos

We will start with a small list of entities using an example, in Table 2, of absolute gains and relative gains of the number of daily views for the trending YouTube videos from October 29 to October 30[th], 2021.

Such lists are regularly published by YouTube.

These videos are ranked by the total number of views on October 29th at 8 pm (the total number is not shown here, but ranges from 128 million views for Adele through tens of millions of views on the top of the list to a few hundred thousand of views – all gained within several days since launch).

Instead, we show here two functions: g (absolute gain) and r (Relative gain) within 24 hours.

| Rank | Name | AbsoluteGain | RelativeGain |
|---|---|---|---|
| 1 | Adele - Easy On Me | 2,752,518 | 2.20% |
| 2 | DJ Snake, Ozuna, Megan Thee Stallion, LISA of BLACKPINK – SG | 2,278,881 | 4.60% |
| 3 | Nardo Wick - Who Want Smoke?? ft. Lil Durk, 21 Savage & G Herbo | 1,989,507 | 4.50% |
| 4 | Anuel AA – Dictadura | 6,301,433 | 33.20% |
| 5 | Olivia Rodrigo – traitor | 505,000 | 6.10% |
| 6 | Lil Durk - Pissed Me Off | 370,806 | 3.30% |
| 7 | Falling (Original Song: Harry Styles) by JK of BTS | 2,018,381 | 19.20% |
| 8 | Swedish House Mafia and The Weeknd | 491,147 | 5.10% |
| 9 | El Alfa "El Jefe" x Dowba Montana x MarkB - Tamo En Hoja | 403,187 | 7.80% |
| 10 | Lil Tjay - Not In The Mood | 350,160 | 7.40% |
| 11 | MOUNT WESTMORE – Big Subwoofer | 317,125 | 7.70% |
| 12 | Moneybag Yo, Lil Durk, EST Gee –Switches & Dracs | 63,112 | 9.50% |
| 13 | 42 Dugg - FREE RIC | 115,024 | 2.80% |
| 14 | Fuerza Regida – Descansando | 180,665 | 14.30% |
| 15 | Kodak Black Ft. Chief Keef - Who Want Smoke Remix | 53,034 | 1.60% |
| 16 | KayyKilo X DaBaby - Yeah B*tch | 103,045 | 3.50% |
| 17 | Jack Harlow - Luv Is Dro | 130,129 | 6.10% |
| 18 | NoCap - Unwanted Lifestyle | 195,417 | 10.70% |
| 19 | Summer Walker - Ex For A Reason (ft. JT From City Girls) | 120,017 | 7.10% |
| 20 | Moneybagg Yo – Scorpio | 63,112 | 9.50% |
| 21 | Icewear Vezzo x Babyface Ray- Sippin | 56,112 | 12.10% |
| 22 | Big Sean, Hit-Boy - Loyal To A Fault ft. Bryson Tiller, Lil Durk | 471,034 | 101.00% |
| 23 | Quando Rondo - Time Spent | 68,105 | 16.30% |
| 24 | Kevin Gates x Dusa "Dear God | 79,078 | 19.10% |
| 25 | Lil Uzi Vert - Demon High | 83,197 | 19.00% |
| 26 | Topher - Let's Go Brandon | 88,105 | 32.00% |
| 27 | Mo3 - Soul Ties (Official Video) ft. Derez De'Shon | 90,567 | 51.00% |
| 28 | Fredo Bang - Many Men (Feat. JayDaYoungan) | 191,004 | 126.00% |

**Table 2.** *Top Trending videos on YouTube (October 28th, 2021) ranked by total number of views (not displayed) and their absolute and relative gain in October 28th (within 24 hours).*

It is easy to see that there are only 3 momentum leaders (Pareto frontier)

These are:

i. *Anuel AA Dictadura* (#4)
ii. *Big Sean* (#22)
iii. *Fredo Bang* (#28)

Indeed, none of the above videos is dominated by another video from the above list.

As expected relative gains become much bigger as we move down the list. Notice that *Fredo Bang* (#28) dominates five music videos between #22 and #28 (22-28). The *Big Sean* (#22) dominates all videos ranked between #7 and #28. *Anuel AA Dictadura* dominates all videos from #1 to #22. All in all, *Anuel AA Dictadura* is the most impressive.. Notice that *Anuel AA Dictadura* has gained over 6 million views (almost 1/3 of its total views) in 24 hours!

***Notice that any entity is dominated by one or more momentum leaders (or is a momentum leader itself)***

The number of such momentum leaders depends on the ∆-system E. In general, the number of momentum leaders (cardinality of Pareto frontier) can vary from just one momentum leader to as many as all entities in E.

This means we face an important practical constraint. While having just one momentum leader is quite enticing, having too many of them loses its value. Momentum should be a rare distinction, not a "common" property, to be a valuable insight for the consumer of the data.

If absolute and relative gains follow a power law, we show the small frontier property - stating that the number of momentum leaders is very small. especially if their scores follow the power law (see further discussion of the role power-law plays in our applications). Using the prior example of YouTube videos, it would be one video that would have both absolute gains in the number of views and relative gain higher than any other video. This would probably have to be a popular video which is still in the process of "taking off", say one with tens or even hundreds of millions of views, with a relative gain of several orders of magnitude (since it would need to have higher relative gain than even videos with very small viewership). Thus, such video would need effectively to have absolute gain in billions, if not tens of billions of views!

The opposite situation would arise when each video is a momentum leader. For example when the relative gain is monotonically increasing for i=1…N. The lower the rank of the entity, the higher the relative gain. In the same time the lower the rank of the entity the lower the absolute gain as well.

Here is some such example E=<N,g,r>:

For entities i=1…N

$g(i) = C/i$

$r(i) = \log(i)$

Notice that for no pair i and j,  i $<<_m$ j.

Indeed, no entity dominates another entity on *both* absolute and relative gain for E.

In practice, with the power-law distribution of scores, absolute gains, and relative gains, we observe a surprisingly small number of momentum leaders, even less than ten for hundreds of thousands of entities! We will illustrate it with our running example, based on the real data set from Articker.

Additionally, we also provide natural ordering between momentum leaders. This can provide a ranking principle to select the top momentum leaders, especially if there are too many of them, This ranking is described below.

### 3.1 Linear ranking of Momentum Leaders

Is there a natural way to further rank the momentum leaders among themselves? Clearly $<<_m$ is not sufficient, since momentum leaders are not comparable.

There is a natural way to order momentum leaders, which we explain below:

Given a momentum leader m, we can define a set D(m) of all entities d such that d $<<_m$ m, that is entities dominated by m. Notice that D(m) and D(m') for two different entities m and m' need not be disjoint.

We can calculate the weight of D(m), w(m), as the sum of scores of all elements of D(m). We can further normalize this weight by dividing it by the union of scores of all entities in E (the Total). In this way can obtain normalized weight w*(e), which is a ratio between 0 and 1. The higher the normalized weight of D(m) is, the more dominant momentum leader m is.

The formula to compute the weight of m is described below

(*)   w(m) =  ∑ w* (e) : e in D(m)

***We propose to rank momentum leaders in descending order of values of w(m)***

D(m) is also characterized by a distribution of scores of members of D(m). We distinguish a very special subset of D(m), called an *Interval(m),* which is the maximal contiguous interval of entities containing m such that Interval(m) is a subset of D(m). These are entities in D(m) which are ranked strictly "around" m, some of them ranked lower than m, others, ranked higher than m. All these entities have lower absolute and relative gain than m.

**Definition** *Interval(m)*

The interval of m is formally defined as follows:

 Interval(m) = <L, R> such that L < R and such that

i.  for each entity n such that L< n and n <R, n is in D(m).
ii. The interval Interval(m) is maximal with such property – for no L'<L or R'>R, <L',R'> has the property (i.).

Thus, Interval(m) is the maximal set of entities that are ranked strictly around m in the form of a contiguous interval, which is dominated by m.

 One may characterize these entities as at least belonging to the same "cohort" as m and carrying less momentum than m. All entities in Interval(m) which are ranked higher than m, are expected to have higher absolute gain than m. But they do not. And all entities in Interval(m) which are ranked lower than m, are expected to have higher relative gain than m. But they don't. This is why m is a momentum leader. It dominates all its peers.

### 3.2 Small Frontier Property

Here we describe how we estimate the size of the set of momentum leaders – the Pareto frontier of $<<_m$

We need the following notion of *moving maximum.*

Definition *Moving Maximums of ordered sequences*.

Let a[1],...a[n] be a sequence of n values. We will call b=a[k]  a moving maximum, if a[k] is larger than all a[i] for i<k.

The number of moving maxima of the relative gain - r[i] provides an important upper bound for the size of the frontier for the delta system E = <N, a,r>. Namely, let M be the number of moving maxima for the sequence r[1],....r[N].

Then the size of Pareto frontier F for E is no larger than M.

**Proof**

We will show that for any entity f = <a[k], r[k]> which is a part of frontier F of momentum leaders, r[k] is a moving maximum for sequence r[1]...r[N].

Assume f is a momentum leader but it is NOT moving maximum. Then there must be g>f such that r(g)> r(f), but then a(g) >a(f) because g>f. But then g>>f and f is not a momentum leader. Contradiction.

Below we present two examples of Relative gain distribution over the ranking of entities in E. The graph below, Figure 2, illustrates circumstances that lead to a small frontier – when the set of moving maxima is small. It happens when moving maxima are "outliers" - they are more than local maxima they are like "record high" values of relative gain for all rankings above their ranking. For example, the entities ranked between M4 and M5, have some local maxima of relative gain, but they are not breaking the record high of M4. This record holds until the entity M5, whose relative gain beats the previous record belonging to M4. In real data, the distance between the successive record high values of relative gain grows exponentially. This is consequence of power law distribution – since it is less and less likely to "break a record" of the highest relative gain as we move to the higher values of relative gain. The growing distance between moving maxima, results in having few of them and consequently, small frontier size.

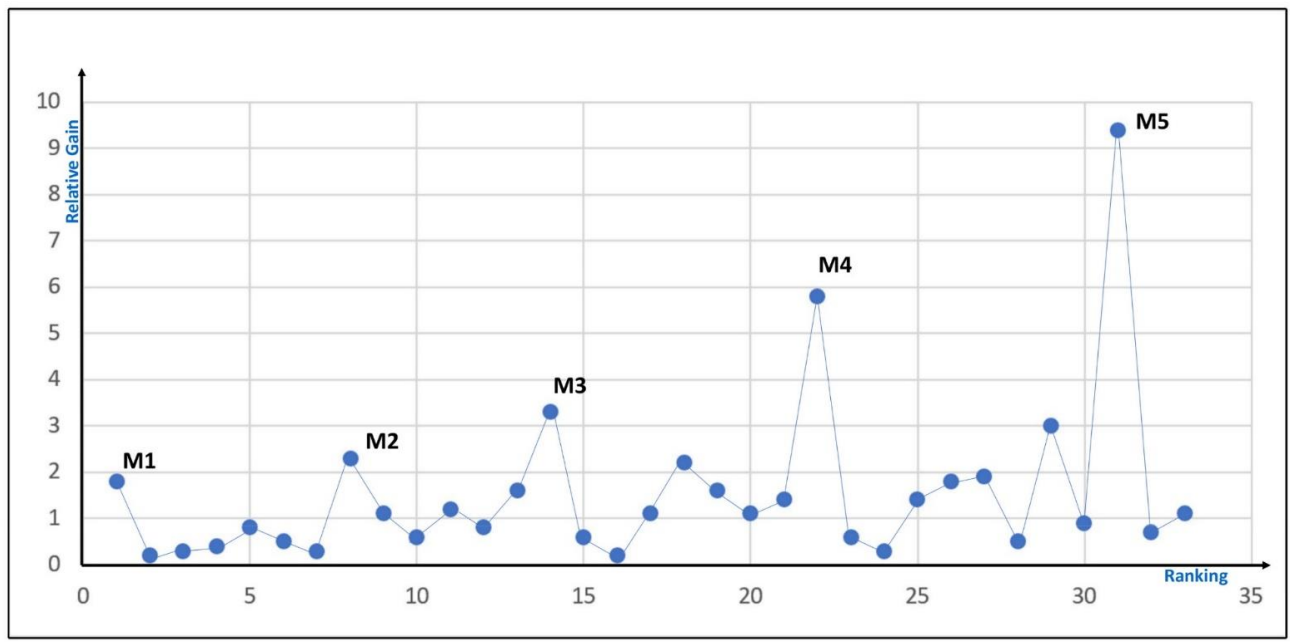

**Figure 2.** *Very few momentum leaders (M1-M5).*

The opposite situation is illustrated below, in Figure 3. The relative gain of entities is monotonically increasing with decreasing ranking. Thus, every single entity posts a new record high relative gain, as compared with entities that are ranked higher. Here the number of moving maxima is equal to the number of all entities.

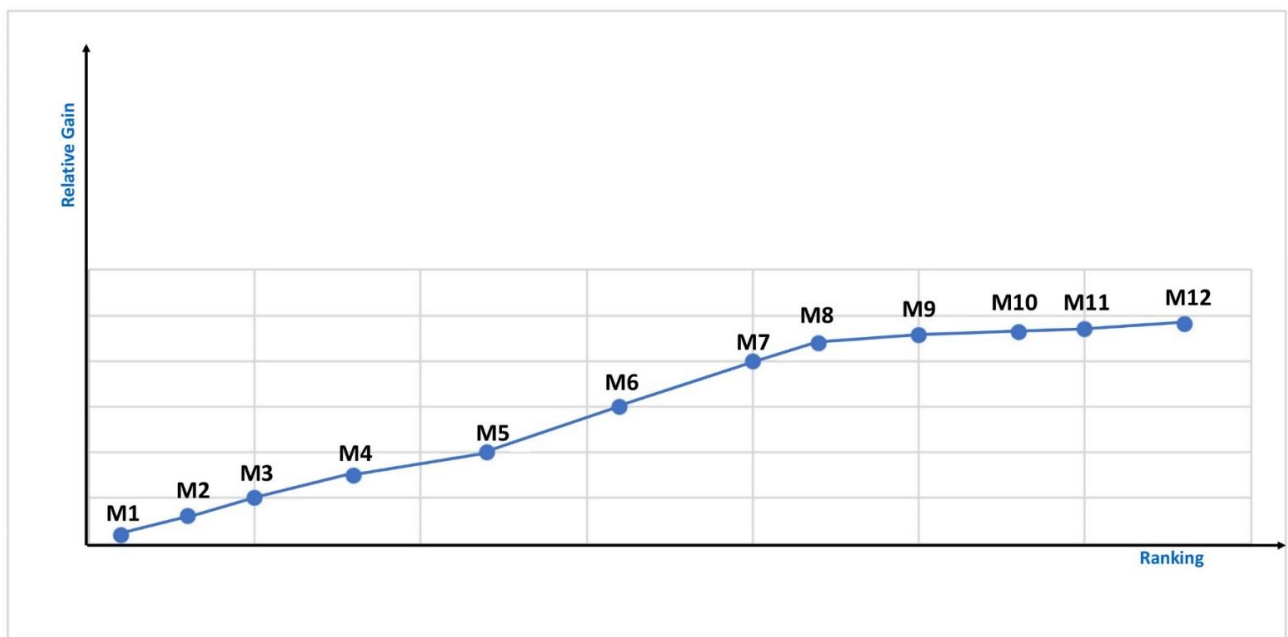

**Figure 3.** *Every entity is a momentum leader (size of Pareto Frontier same as the number of all entities).*

Notice that relative gains of entities are slowly growing as a function of ranking of the entities by score, the lower the rank, the higher relative gain. There is no momentum here – since all entities are changing "in tandem" with relative gain growing slightly in synchronization with a lower rank.

The data sets with power-law distributions of g and r are much closer to Figure 3. Very few moving maxima, leading to small Pareto frontier.

In general, we have observed that when the distribution of absolute and relative gains follow power-law then the size of the frontier is very small, several orders of magnitude smaller than the total number of entities. This is really good news – since we want the set of momentum leaders to be selective and very small in comparison to the set of all entities. We have run 100,000 Monte Carlo simulations systems with N >20,000 and have established that the size of the frontier for power-law distributions is bounded by the square of logarithm of the size of the entire entity set c *(log(N)+1)^2* where c is a constant between 0.3 and 0.5 depending on the top percentile (from 95[th] to 99[th] percentile of size values)

### 3.3 Monte Carlo Simulations

We have tested 100 simulated Δ-systems of different sizes ranging from 20,000 to 200,000 entities, and for each of them we took the distributions of values of r and g and permuted them 1000 times. We computed a boxplot of the size of the Pareto frontier and took the upper bound for 95 and 99 percentile of Pareto frontier size, Table 3.

| Frontier Size | Numbers of Entities | Percentile |
|:---:|:---:|:---:|
| 9.30 | 20,000 | 95 |
| 14.10 | 20,000 | 99 |
| 10.80 | 50,000 | 95 |
| 16.20 | 50,000 | 99 |
| 12.00 | 100,000 | 95 |
| 18.00 | 100,000 | 99 |
| 13.20 | 200,000 | 95 |
| 19.80 | 200,000 | 99 |

**Table 3.** *Frontier size for different number of entities (Monte Carlo simulations).*

In general, for sufficiently large N, the size of the frontier can be estimated between $(1/3) * (\log(N)+1)^2$ and $(1/2) * \log(N)+1)^2$ depending on the percentile of 95 or 99.

These simulations assume power-law distributions of absolute and relative gains with the exponent of -1.

The informal explanation has to do with the small number of very large values of relative gains and absolute gains (for opposite ends of the ordinal values). Since we use the number of moving maxima as the upper bound of the frontier size, let us phrase our explanation in terms of the relative gains sequence. The "outliers" of relative gain appear more and more often for entities ranked lower and lower. They form moving maxima which dominate longer and longer intervals, resulting in exponentially growing distance between successive moving maxima.

### 4. Applications Examples

We will start with very small sets in widely popular domains of music and stocks. We just would like to show entities that are selected as momentum leaders by our algorithm, rather than demonstrate the sizes of frontiers (our estimates apply to much larger entity sets). The latter we will demonstrate with Articker data set as the last of applications which we describe here.

### 4.1 YouTube music videos

The list below, in Table 4, contains the top 20 pop music videos ranked by the total number of views (this list excludes the nursery song videos – some of which have acquired billions of views). While we do not show the total number of views (ranging between 7 billion down to around 3 billion), we display the AbsoluteGain and RelativeGain for one day, October 28[th], 2021. For example, Luis Fonsi, who tops the list, has gained almost 1.7M views on that day (notice that it is less than some of the trending videos! but videos on this last have been on youtube in some cases, several years). Relative Gains are much smaller, as expected. They range around a small fraction of a percent.

There is *only one* momentum leader for this list and it is number 6, *El Chombo*. He dominates all other artists in Pareto ordering of both Absolute and Relative Gains (Both absolute gain of 2,287,210 and relative gain of 0.062% are the highest among this list of 20.

Thus, this list has a singular Pareto frontier.

One could think about an interesting notion of the runner(s) up – if we removed El Chombo from the list (hypothetically), what would be the new momentum leaders? It would be number 1, Luis Fonsi, number 2, Ed Sheehan, number 5, PSY, and number 20, another music video by Ed Sheehan.

| Rank | Name | AbsoluteGain | RelativeGain |
|---|---|---|---|
| 1 | Luis Fonsi - Despacito ft. Daddy Yankee | 1,699,966 | 0.022% |
| 2 | Ed Sheeran - Shape of You | 1,300,836 | 0.023% |
| 3 | Wiz Khalifa - See You Again ft. Charlie Puth | 1,197,170 | 0.022% |
| 4 | Mark Ronson - Uptown Funk | 812,456 | 0.018% |
| 5 | PSY – Gangnam Style | 1,216,920 | 0.028% |
| 6 | El Chombo - Dame Tu Cosita feat. Cutty Ranks | 2,287,210 | 0.062% |
| 7 | Maroon 5 - Sugar | 753,701 | 0.021% |
| 8 | Justin Bieber - Sorry | 314,348 | 0.011% |
| 9 | Katy Perry - Roar | 772,098 | 0.021% |
| 10 | OneRepublic - Counting Stars | 883,082 | 0.023% |
| 11 | Ed Sheeran - Thinking Out Loud | 643,832 | 0.018% |
| 12 | Alan Walker - Faded | 660,254 | 0.019% |
| 13 | Katy Perry - Dark Horse | 649,985 | 0.019% |
| 14 | Maroon 5 - Girls Like You ft. Cardi B | 909,537 | 0.028% |
| 15 | Major Lazer & DJ Snake - Lean On | 471,952 | 0.015% |
| 16 | Enrique Iglesias - Bailando ft. Descemer Bueno, Gente De Zona | 559,259 | 0.016% |
| 17 | Taylor Swift - Shake It Off | 358,468 | 0.012% |
| 18 | Passenger - Let Her Go | 714,172 | 0.023% |
| 19 | Balvin, Willy William - Mi Gente | 447,575 | 0.016% |
| 20 | Ed Sheeran - Perfect | 1,090,884 | 0.036% |

**Table 4.** *Top 20 music videos ranked by total number of views (not shown) and their absolute and relative gain of views in one day (October 28th 2021).*

### 4.2 Stocks

We have looked at two different lists for October 28th, 2021. One was the top 20 market cap stocks and their gains and losses during that Friday.

Which stocks were momentum leaders?

| Rank | Name | Symbol | Marketcap | Price(USD) | Country | RelativeGain |
|---|---|---|---|---|---|---|
| 1 | Microsoft | MSFT | 2.49E+12 | 331.62 | United State | 2.24% |
| 2 | Apple | AAPL | 2.28E+12 | 149.8 | United States | -1.82% |
| 3 | Saudi Aramaco | 2222.SR | 2.01E+12 | 10.06 | Saudi Arabia | 0.13% |
| 4 | Alphabet (Google) | GOOG | 1.97E+12 | 2965.41 | United States | 1.47% |
| 5 | Amazon | AMZN | 1.71E+12 | 3372.43 | United States | -2.15% |
| 6 | Tesla | TSLA | 1.12E+12 | 1114 | United States | 3.43% |
| 7 | Facebook | FB | 9.00E+11 | 323.57 | United States | 2.10% |
| 8 | Berkshire Hathaway | BRK-A | 6.51E+11 | 432902 | United States | -0.77% |
| 9 | NVIDIA | NVDA | 6.37E+11 | 255.67 | United States | 2.51% |
| 10 | TSMC | TSM | 5.90E+11 | 113.7 | Taiwan | -1.99% |
| 11 | Tencent | TCEHY | 5.83E+11 | 60.79 | China | -3.08% |
| 12 | JPMorgan Chase | JPM | 5.02E+11 | 169.89 | United States | -0.28% |
| 13 | Visa | V | 4.62E+11 | 211.77 | United States | 0.91% |
| 14 | Alibaba | BABA | 4.47E+11 | 164.94 | China | -2.86% |
| 15 | UnitedHealth | UNH | 4.34E+11 | 460.47 | United States | 1.10% |
| 16 | Johnson & Johnson | JNJ | 4.29E+11 | 162.88 | United States | 0.02% |
| 17 | Walmart | WMT | 4.17E+11 | 149.42 | United States | 0.65% |
| 18 | Samsung | 005930.KS | 4.02E+11 | 59.72 | South Korea | -1.27% |
| 19 | LVMH | LVMUY | 3.96E+11 | 157.36 | France | 0.76% |
| 20 | Bank of America | BAC | 3.94E+11 | 47.78 | United States | 0.00% |

**Table 5.** *Top 20 stocks ranked by market capitaization (Marketcap) and their relative gain on October 28th.*

These stocks, in Table 5, have been ranked by market capitalization. The absolute gains can be easily computed by multiplying the market caps of the stocks by Gain (it is Relative Gain). For this list, we have just two momentum leaders – Microsoft (ranked number 1 by market cap) and Tesla (ranked number 6). Indeed, Microsoft dominates all stocks down to number 6, Tesla. Tesla, on the other hand, dominates all stocks except Microsoft. Thus, Tesla is the overall weighted momentum leader for the top 20 by Market cap, Friday, October 28th list. Notice that any stock is dominated by either Tesla or Microsoft *or both*. Also, if Tesla gained more than 6%, it would be the *sole momentum leader*, dominating Microsoft both on absolute gain (market cap times relative gain) as well as on the relative gain.

Finally, we have obtained the list of stocks that were top gainers relatively on Friday, October 28th, 2021. These are stocks, in Table 6, with much smaller market caps than the list of top 20, nevertheless few of them exceeded $1 Billion market cap, and one (ranked number 1 by market cap), QuantumScape has a market cap of over $12 Billion.

| Rank | Name | Symbol | AbsoluteGain | RelativeGain |
|---|---|---|---|---|
| 1 | QuantumScape | QS | 1756 | 13.94% |
| 2 | Matterport | MTTR | 1004 | 18.66% |
| 3 | Live Oak Bancshares | LOB | 563 | 14.60% |
| 4 | Cadence Bank | CADE | 1329 | 44.38% |
| 5 | Arqit Quantum | ARQQ | 567 | 21.87% |
| 6 | Lending Tree | TREE | 305 | 14.27% |
| 7 | Aspen Aerogels | ASPN | 320 | 17.83% |
| 8 | Berkshire Grey | BGRY | 285 | 8.15% |
| 9 | A10 Networks | ATEN | 499 | 34.96% |
| 10 | Vicarious Surgical | RBOT | 204 | 15.87% |
| 11 | GH Research PLC | GHRS | 283 | 22.92% |
| 12 | Bakkt Holdings | BKKT | 582 | 66.09% |
| 13 | Ramaco Resources | METC | 141 | 16.99% |
| 14 | Bit Digital | BTBT | 328 | 41.79% |
| 15 | Arteris | AIP | 161 | 23.75% |
| 16 | MiNK Therapeutics | INKT | 140 | 23.29% |
| 17 | Decarbonization Plus Acquisition | DCRCU | 89 | 17.70% |
| 18 | Plx Pharma | PLXP | 49 | 14.50% |
| 19 | Eqonex Limited | EQOS | 120 | 47.31% |
| 20 | Remark Holdings | MARK | 31 | 14.05% |

**Table 6.** *Top 20 stocks with highest relative daily gain on October 28th, 2021. Stocks are ranked by market capitalization (not shown). Absolute daily gain (in market cap) and relative gain shown.*

What are the momentum leaders on this list? Number 1, Quantum Scape, number 4, Cadence Bank, and number 12, Bakkt Holdings are momentum leaders as they are not dominated by any stock on this list. Any other stock on this list is dominated in Pareto Ordering by one of these three momentum leaders.

If we applied the weighted linear ordering of these stocks, Cadence Bank (number 4) dominates all entities 2,3, 5…11. We have to calculate the total weight (in terms of market cap share) of this set and it was larger than the market share D(m) for remaining momentum leaders.– Notice also the interval intuition – CADE dominates its "cohort" of stocks that are compatible with CADE on this list – stocks with closer market cap to CADE's market cap. With 44.38% relative gain it dominated all its cohort members on that Friday, October 28th.

### 4.3 Artists

The last domain is the domain of the visual artist – the subject of our platform Articker. This is where our motivation to create ranking by momentum came from. This list has over 100,000 artists and every month we use our Ranking by Momentum algorithm to publish momentum leaders for the past 90 days on www.articker.org . We describe the Articker application in much more detail in the companion paper [2].

Here, as a simple example, in Table 7, we just look at the top 20 artists on the list according to the media index (an Articker defined metric, which is proportional to the media presence of an artist). One can see many well-recognized names – leading with Picasso and Warhol. The media index is moving much slower than views of music videos or as stocks. Therefore we look here not at one day, like in the case of videos and stocks, but at the period of 90 days. The Total gain shows the media index increase over the last 90 days (This list was constructed for August 15, 2021, so it is pretty recent). The relative gain is computed as the gain or loss of media index share for each artist. This notion is analogous to the market share – and divides the media index value of a given artist by the total media index of all artists at any point in time. For example relative gain of Picasso of 0.18, means that he gained 0.18% of media share over 90 days ending in August. Since Picasso has such dominating overall lead in the media index (with 490,000 points, he doubles Warhol's who is second), the positive relative gain is quite impressive.

| Id | Name | Totalgain | Relativegain |
|---|---|---|---|
| 1 | Picasso | 13076.67 | 0.18% |
| 2 | Andy Warhol | 6534.81 | -0.22% |
| 3 | Van Gogh | 10649.86 | 2.15% |
| 4 | Banksy | 11872.99 | 2.52% |
| 5 | Leonardo DaVinci | 2007.49 | -0.98% |
| 6 | Rembrandt | 2353.04 | -0.44% |
| 7 | Damien Hirst | 2618.22 | -0.10% |
| 8 | Ai Weiwei | 1551.21 | -1.06% |
| 9 | Basquiat | 5355.81 | 3.02% |
| 10 | Jeff Koons | 2474.82 | 0.08% |
| 11 | David Hockney | 2091.54 | -0.18% |
| 12 | Claude Monet | 2278.12 | 0.31% |
| 13 | Salvador Dali | 1963.00 | -0.02% |
| 14 | Caravaggio | 2480.67 | 0.69% |
| 15 | Yayoi Kusama | 2102.74 | 0.24% |
| 16 | KAWS | 1153.46 | 1.57% |
| 16 | Francis Bacon | 1907.08 | 0.11% |
| 17 | Qi Baishi | 198.22 | -2.18% |
| 18 | Frida Kahlo | 3190.99 | 2.14% |
| 19 | Jackson Pollock | 1132.58 | -0.80% |

Table 7. *Top 20 artists ranked by overall Articker media index value Total gain over 90 days preceding October 1, 2021 as well as relative gain of media index share over the 90 day window displayed. Notice – here relative gain can be negative – since it reflects the change in over media index share for the given artist.*

Who are the momentum leaders for this top 20 list?

Picasso is one of them, so are Van Gogh and Banksy. Indeed Picasso is not dominated by any artist, but he does not dominate Van Gogh (who has higher relative gain), Van Gogh is not dominated by anyone, but does not dominate Banksy. Both Banksy and Van Gogh dominate every artist down the list until Basquiat with his 3.02% relative gain. Basquiat dominates all artists ranked below him (he leads them both by absolute gain and relative gain), he also dominates artists ranked higher by him: DaVinci, Rembrandt, Damien Hirst, and Ai Weiwei. These artists are also dominated by Banksy, Van Gogh, and also by Picasso.

Thus, we have 4 momentum leaders here- Picasso, Van Gogh, Banksy, and Basquiat.

Who has the highest weight as computed by (*)?

Turns out that it is Basquiat, followed by Van Gogh, Banksy, and Picasso. This is of course a "battle of giants" among elite blue-chip artists. In [2] we describe how momentum ranking helps us to establish momentum leaders among *all* artists

## 5. Comparing two Δ-systems about how momentous they are?

So far we have investigated finding entities with the most momentum for a given system delta system E.

Different systems may carry different momentum as a whole.

This leads us to a question:

How "momentous" is a Δ-system E?

The higher the relative gain of the total score of all entities in E, the more momentous is E. If the total score of all entities goes up by 50%, it is more momentous than if it grew only by 10%. But it is not the whole story. What matters is how many and how big are the outliers in each case as compared to the "base" relative gain.

Let us remind that w(m) is the relative total weight of D(m), given a momentum leader m.

The following formula describes the momentum of Δ-system E:

$$m(E^*) = \sum r(m) * w(m)$$

where the summation is over all momentum leaders m of E*

The momentum of E rewards high relative gains of momentum leaders m who at the same time dominate entity sets of high relative weight.

This way we can compare two different systems E and F and decide which one is more momentous. We may develop a scale of *momentousness* and decide how *momentous* a given system is. This is particularly useful when publishing periodic lists of top entities with the highest momentum (say monthly lists).

Is August's list of artists more momentous (interesting) than the July list? Is the stock list of gainers for last Monday, more momentous than the list for last Tuesday? Is the list of trending videos for last Saturday more momentous than the one from one week earlier? It depends on the outliers among entities and how far out in terms of relative gain they were. The more extreme outliers they are, the more momentous and interesting the system is.

### Example

Let E*= <N, g1,r1> consists of 10,000 entities and F*=<M,g2,r2> consist of 100,000 entities. Let us assume that E* has 3 momentum leaders k1,k2, k3, and F* has 5 momentum leaders l1, l2,l3,l4,l5

Assume that all entities in E* gain an overall 10% and all entities in F only gain 2% overall. Nevertheless, if F* has one huge outlier it may be more momentous than E* as the following tables show:

| Momentum Leader | Relative Gain | Weight |
|---|---|---|
| k1 | 20% | 10% |
| k2 | 40% | 60% |
| k3 | 80% | 30% |

Table 8. *Δ-system with no relative gain outliers.*

Overall momentum of E* = 0.2 * 0.1 + 0.4*0.6 + 0.8 *0.3 = 0.50

| Momentum Leader | Relative Gain | Weight |
|---|---|---|
| l1 | 5% | 10% |
| l2 | 10% | 30% |
| l3 | 20% | 20% |
| l4 | 50% | 20% |
| l5 | 1000% | 20% |

Table 9. *Δ-system with significant relative gain outlier.*

Overall momentum of F* = 0.05 * 0.1 + 0.1*0.3 + 0.2*0.2 + 0.5*0.1 + 10* 0.2 = 2.125

Due to the presence of the very high relative gain momentum leader, l5 (1000%), F* is much more momentous than E*, quite independently of the overall rate of change.

## 6. Conclusions

The notion of momentum is fundamental and applies to a wide range of applications when a large number of entities change over time. We would like to identify entities with momentum, that is entities that are changing not just the most rapidly but also with the most gravitas. The problem is inherently two-dimensional, just like physical momentum which depends on the mass and velocity of the moving object. Absolute gain and relative gain are two dimensions which for an individual entity are of course dependent on each other (one can be calculated from another), but are characterized by almost opposite distributions. Relative gain is larger with the lower ranking of an entity by score. Thus it is negatively correlated with the score, the lower the score, the higher on average, relative gain. On the other hand, the absolute gain is positively correlated with the score.

It turns out that Pareto Ordering $>>_m$ on both absolute gain and relative gain provides the right vehicle for ranking by momentum. Pareto frontier for $>>_m$ as a vehicle to determine momentum leaders in this two-dimensional environment. We showed experimentally that if absolute gain and relative gain are distributed according to power law then Pareto frontier size is very small, of the order of $\log(N)^2$. We call it small Pareto frontier property.

We demonstrate how our algorithm to determine momentum leaders work for various domains such as YouTube music video ranking, stock ranking as well as Articker ranking of an artist – which originally motivated this work.